\documentclass[a4paper,10pt,onecolumn,oneside,notitlepage,openbib]{article}
\usepackage{amssymb}
\usepackage{amsmath}
\usepackage{latexsym}
\usepackage{array}
\usepackage{graphics}

\def\DIS{\displaystyle}
\def\hbreak{\vspace*{3mm}\hfill\break\noindent}

\def\LCM{\mbox{L.C.M.}}
\def\GCD{\mbox{G.C.D.}}
\def\qed{\hfill\hbox{$\Box$}\vspace{10pt}\break}
\usepackage{theorem}
\theoremstyle{break}
\newtheorem{Theorem}{Theorem}[section]
\def\Proof{\hfil\break{\bf Proof}}
\newtheorem{Proposition}{Proposition}[section]
\newtheorem{Lemma}{Lemma}[section]
\newtheorem{Corollary}{Corollary}[section]

\newtheorem{Definition}{Definition}[section]
\numberwithin{equation}{section}

\def\Z{{\mathbb Z}}
\def\Q{{\mathbb Q}}

\def\mapto{\rightarrow}

\author{
Daisuke Yoshihara\thanks{Graduate school of Mathematical Sciences,
      University of Tokyo, 3-8-1 Komaba, Tokyo 153-8914, Japan}, \ 
 Fumitaka Yura\thanks{Imai quantum computing and information project,
    ERATO, JST, Daini Hongo White Bldg. 201, 5-28-3 Hongo, Bunkyo,
    Tokyo 113-0033, Japan} \ 
and Tetsuji Tokihiro$^*$
}
\title{Fundamental Cycle of a Periodic Box-Ball System}
\date{}

\begin{document}
\maketitle

\begin{abstract}
We investigate a soliton cellular automaton (Box-Ball system)
 with periodic boundary conditions.
Since the cellular automaton is a deterministic dynamical system 
that takes only a finite number of states, it will exhibit periodic motion. 
We determine its fundamental cycle for a given initial state.
\end{abstract}

\section{Preface}

A cellular automaton (CA) is a discrete dynamical system consisting of
a regular array of cells\cite{Wolfram}.
Each cell takes only a finite number of states and is updated 
in discrete time steps.
Although the updating rules are simple, CAs often exhibit very complicated 
time evolution patterns which resemble natural phenomena such as chemical
 reactions, turbulent flow, nonlinear dispersive waves and solitons.
A typical CA exhibiting a solitonic behaviour is the box and ball system (BBS) 
which is a reinterpretation of the CA proposed by Takahashi and 
Satsuma\cite{TS,Takahashi2}.
The BBS is {\it integrable} in the sense that it is obtained from the KdV 
equation through a limiting procedure called ultradiscretization\cite{TTMS}. 
It can also be obtained from a two dimensional integrable lattice model 
and its relation to combinatorial $R$ matrices of ${U'}_q(A_N^{(1)})$ 
is well established\cite{FOY, HHIKTT}.

The original box-ball system is defined as a dynamical system of a finite 
number of balls in an {\it infinite} one dimensional array of boxes. 
However, it is possible to extend the time evolution rule to a system 
consisting of a
{\it finite} number of boxes with periodic boundary conditions\cite{YT}. 
The BBS with periodic boundary condition (pBBS) is also connected to the 
combinatorial $R$ matrix of ${U'}_q(A_N^{(1)})$ and, its time evolution rule 
is represented as a Boolean recurrence formula related to the algorithm for 
calculating the $2N$th root. 
As a CA, the pBBS is composed of a finite number of cells, 
and it can only take on a
finite number of patterns. 
Hence the time evolution of the pBBS is necessarily periodic.
In the present article, we investigate the fundamental cycle, {\it i.e.}, 
the shortest period of the discrete periodic motion of the pBBS.\footnote{
Part of the present work was already announced in Ref.\cite{Yoshihara}.}
In section 2, we review the pBBS and present its conserved quantities.
The formula for the total number of patterns for a
 given set of conserved quantities is also presented.
In section 3, we define several notions which are necessary to prove the 
formula for the fundamental cycle.
The fundamental cycle for an arbitrary initial state is derived in section 4. 
Section 5 is devoted to the concluding remarks.
%
%
\section{Periodic box and ball system (pBBS)}

\begin{figure}[!b]
  \begin{center}
     \scalebox{0.45}{\includegraphics{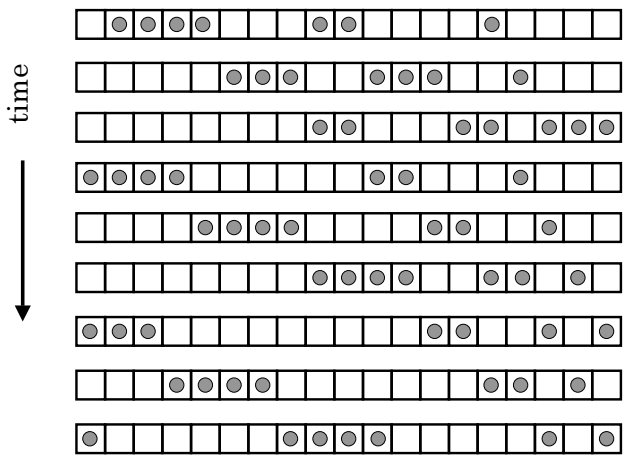}}
  \end{center}
  \caption{Time evolution of a pBBS.}
  \label{fig:pBBS-1}
\end{figure}

\begin{figure}[htbp]
  \begin{center}
     \scalebox{0.7}{\includegraphics{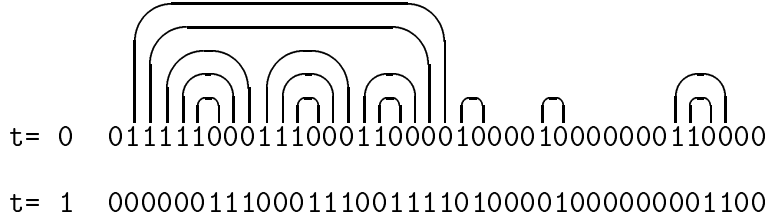}}
  \end{center}
  \vspace{-5mm}
  \caption{Time evolution rule for a pBBS expressed as $10$ sequences.}
  \label{fig:pBBS-2}
\end{figure}

\begin{figure}[!b]
  \begin{center}
     \scalebox{0.7}{\includegraphics{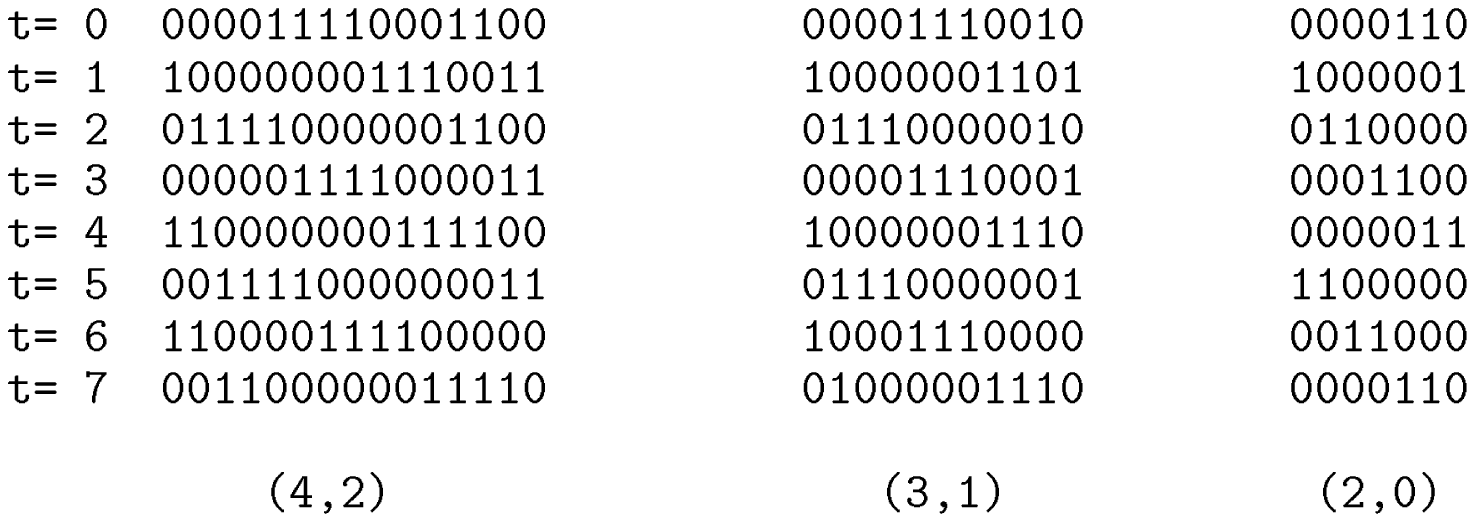}}
  \end{center}
  \vspace{-5mm}
  \caption{Examples of time evolution patterns of pBBSs. 
The middle (right) pBBS is obtained fron the left (middle)
 pBBS by $10$ elimination. 
The numbers in the parentheses denote lengths of {\it solitons} 
in the patterns.}
  \label{fig:pBBS-3}
\end{figure}

Let us consider a one dimensional array of $N$ boxes.
To be able to impose a periodic boundary condition, we assume that
the $N$th box is the adjacent box to the first one. 
(We may imagine
that the boxes are arranged in a circle.)
The box capacity is one for all the boxes, and each box is 
either empty or filled with a ball at any time step.
We denote the number of balls by $M$, such that
$\DIS M < \frac{N}{2}$.
The balls are moved according to a deterministic time evolution rule.
There are several equivalent ways to describe this rule.
For example,
\begin{enumerate}
  \item In each filled box, create a copy of the ball.
  \item Move all the copies once according to the following rules.
  \item Choose one of the copies and move it to the nearest 
        empty box on the right of it.
  \item Choose one of the remaining copies and move it to the nearest  
        empty box on the right of it.
  \item Repeat the above procedure until all the copies have moved.
  \item Delete all the original balls.
\end{enumerate}
An example of the time evolution of the pBBS according to this rule is shown in Fig.~\ref{fig:pBBS-1}.
It is not difficult to prove that the obtained result does not depend 
on the choice of 
the copies at each stage and that it coincides with the evolution rule of the 
original BBS when $N$ goes to infinity.
An advantage of the above description is that we can easily extend the evolution rule to the system with many kinds of balls and many kinds of box capacities.
In the present article, however, we restrict ourselves to the pBBS with only one kind of ball and with box capacity one, and we revert to the following 
description which is more 
convenient to determine the fundamental cycle.
We denote an empty box by $'0'$ and a filled box by $'1'$.
Then the pBBS is regarded as a dynamical system of a finite sequence 
of $'0'$s and $'1'$s.
Because of the periodic boundary condition, we consider the last entry 
in the sequence to be adjacent to the first entry.
(In this sense, the sequence should be regarded as a ring which consists of $'0'$s and $'1'$s.)
We say two sequences are {\it equivalent} 
if one coincides with the other by translation.
For example, the sequence $011101100100001$ is equivalent to $101100100001011$.
The updating rule of the sequence is given as:
\begin{enumerate}
  \item Connect all the $'10'$s in the sequence with arc lines.
  \item Neglecting the $10$ pairs which were connected in
        the first step, connect all the remaining $'10's$ with arc lines.
  \item Repeat the above procedure 
        until all the $'1'$s are connected to $'0'$s.
  \item Interchange every $'1'$ and $'0'$ which are connected to each other
        by an arcline.
\end{enumerate}
This rule is illustrated in Fig.~\ref{fig:pBBS-2} and some
examples of time evolutions of the pBBS are shown in Figs.~\ref{fig:pBBS-3}.

Just as the original BBS\cite{TTS}, the pBBS has a number of 
conserved quantities.
Let $p_1$ be the number of the $10$ pairs in a state of the pBBS, 
which are connected with arc lines 
in the first step of the above time evolution rule.
Similarly, we denote by $p_2$ the number of $10$ pairs connected 
in the 2nd step,  by $p_3$ those in the 3rd step, ..., and by $p_l$ 
those in the last step. 
Clearly  $\DIS p_1 \ge p_2 \ge \cdots \ge p_l$. 
For example, in the state shown in Fig.~\ref{fig:pBBS-2}, 
$\DIS p_1=6, \,p_2=4, \,p_3=2, \,p_4=1, \,p_5=1$.
\begin{Proposition}
\label{conservation}
The numbers $p_i$ in the weakly decreasing series of integers 
$\{p_1, p_2, \ldots, p_l\}$ are 
conserved quantities for the time evolution of the pBBS.
\end{Proposition}
To prove Proposition \ref{conservation}, we prepare two Lemmas.
The number of $'10'$s in a state of the pBBS is equal to that of 
$'01'$s, for they coincide with the number of the set of 
consecutive '1's (or '0's). 
According to the evolution rule, a $'10'$ at $t$ turns into a $'01'$ at $t+1$ 
and all the $'01'$s at $t+1$ are created in this way. 
Hence we find that the number of $'10'$s at $t$ is equal to that 
of $'01'$s at $t+1$.
Thus we have the following Lemma:
\begin{Lemma}
\label{Lemma1}
The number of $'10'$s in the pBBS does not change in time.
It also coincides with the number of $'01'$s at each time step.
\end{Lemma}
When we eliminate the $'10'$s in a state, we obtain a new sequence.
Hereafter we call this operation {\it $10$-elimination}.
(In order to associate this new sequence to the state of a pBBS with a smaller 
number of boxes and balls, we have to determine the position of the first 
entry of the sequence. 
We shall consider this problem in the subsequent section however, 
as it is not necessary for the proof considered here.)
The operation {\it $01$-elimination} is defined in a similar way.
By $10$-elimination, all the series of consecutive $'1'$s 
and consecutive $'0'$s 
in the state have their length reduced by one.
The $01$-elimination has the same effect on the state, hence, 
the sequence obtained by $01$-elimination is equivalent to the one 
by $10$-elimination.
When the sequence obtained by $10$-elimination is updated according 
to the time evolution rule, it becomes equivalent to the sequence 
obtained by $01$-elimination on the state in the next time step. 
(This fact is easily understood from Fig.~\ref{fig:pBBS-2}.) 
Thus we have 
\begin{Lemma}
\label{Lemma2}
Both $10$-elimination and $01$-elimination commute with the updating 
rule of the pBBS,
that is, the sequence obtained by $10$-elimination (or $01$-elimination) 
at $t+1$ is equivalent to the one updated from the sequence obtained by $10$ 
elimination (or $01$-elimination) at $t$.
\end{Lemma}

\Proof {\bf \,of Proposition \ref{conservation}} \hspace{12pt}
 From Lemma \ref{Lemma1}, $p_1$ does not change in time. 
Since $p_2$ is the number of $'10'$s in the sequence obtained by 
$10$-elimination, from Lemma \ref{Lemma1} and \ref{Lemma2}, 
it is also a conserved quantity in time. Repeating the $10$-elimination 
and using Lemma \ref{Lemma1} and \ref{Lemma2}, we find that all the $p_j$ 
are conserved in time.
\qed
Since $\{p_1, p_2, \ldots, p_l\}$ is a weakly decreasing series of positive 
integers, we can associate it with a Young diagram with $p_j$ 
boxes in the $j$-th column ($j=1, 2, \ldots, l$).
Then the lengths of the rows are also weakly decreasing positive integers, 
and we denote them
$$
 \{\underbrace{L_1, L_1, \ldots, L_1,}_{\mbox{ $n_1$ }}
\underbrace{L_2, L_2, \ldots, L_2,}_{\mbox{ $n_2$ }}
\cdots,
\underbrace{L_s, L_s, \ldots, L_s}_{\mbox{ $n_s$ }}\}
$$
where $L_1 > L_2 > \cdots > L_s$.
The set $\DIS \{L_j, n_j\}_{j=1}^s$ is an alternative expression of 
the conserved quantities of the system.
Let $\ell_0 := N-\sum_{j=1}^l 2p_j = N-\sum_{j=1}^s 2 n_j L_j$, and
\begin{eqnarray}
\ell_j &:= &L_j-L_{j+1} \qquad (j=1,2,\ldots,s-1)\\
N_j &:= &\ell_0+2 n_1 (L_1-L_{j+1}) + 2 n_2 (L_2-L_{j+1})+...+2 n_j(L_j-L_{j+1}).
\end{eqnarray}
Then, for a fixed number of boxes $N$ and conserved quantities 
$\{L_j, n_j \}$, the number of possible states of the pBBS 
$\Omega(N;\{L_j, n_j \})$ is given by the following Proposition.
\begin{Proposition}
\label{StatesCount}
\begin{eqnarray}
\Omega(N;\{L_j, n_j \}) &= &\frac{N}{\ell_0} {\ell_0+n_1-1 \choose n_1}
{N_1+n_2-1 \choose n_2}{N_2+n_3-1 \choose n_3}\nonumber \\
&&\qquad \times \cdots \times {N_{s-1}+n_s-1 \choose n_s}
\label{NumberofStates}
\end{eqnarray}
\end{Proposition}
Although the Proposition can be proved by elementary combinatorial arguments,
it is convenient to use notions defined 
in subsequent sections and we thus refer the proof to  Appendix A.
The fact, however,  that the right hand side of (\ref{NumberofStates}) is 
an integer is confirmed using the Lemma:
\begin{Lemma}
\label{Lemma3}
Let $\{a_i\}_{i=1}^s, \{b_i\}_{i=1}^s$ be $2s$ positive integers 
and suppose that the $\{a_i\}_{i=1}^s$ do not have a common divisor, 
$i.e.$, $\ \GCD(a_1, a_2, ..., a_s)=1$.
If there exists a $k$ such that
$$
\frac{b_1}{a_1}=\frac{b_2}{a_2}=\cdots =\frac{b_s}{a_s}= k,
$$ 
then $k$ is an integer.
\end{Lemma}
%
%

\section{Soliton in pBBS and its properties}
Let the boxes be numbered $1, 2, 3, \ldots, N$ from left to right.
Accordingly, the position of an entry in the associated $01$ sequence
 is denoted by the number of the corresponding box.
Because of the periodic boundary condition, we always use the convention 
that the numbers are defined in  $\Z_N$, that is, $j \equiv j+N$, and an 
inequality such as $i<j<k$ means that $j \in \{i+1, i+2, \ldots, k-1\}$.
In order to explain some important features of the pBBS, 
we assign an integer index to 
every entry of the sequences obtained at each time step in the updating rule.
Precisely speaking, we define the map 
$\phi_t: \Z_N \mapto \Z \sqcup \{-\infty\} $ in the following way.
\begin{enumerate}
\item Choose the $10$ pairs in the construction of the conserved quantity $p_1$.If the positions of $'1'$s and $'0'$s are $i_1, i_2, \ldots, i_{p_1}$ and $j_1, j_2, \ldots, j_{p_1}$ respectively, then $\phi_t(i_k)=1,\ \phi_t(j_k)=-1$ $(1 \le k \le p_1)$.
\item Next, choose the $10$ pairs for the conserved quantity $p_2$.
If the positions of $'1'$s and $'0'$s are $i_1, i_2, \ldots, i_{p_2}$ and $j_1, j_2, \ldots, j_{p_2}$ respectively, then $\phi_t(i_k)=2,\ \phi_t(j_k)=-2$ $(1 \le k \le p_2)$.
\item Similarly, the indices $3,-3,4,-4,...$ are assigned to the positions 
of the $10$ pairs in the construction of $p_3, p_4, \ldots$.
\item For the positions of $'0'$s which are not connected with $'1'$s in the construction of the conserved quantities, the indices are $-\infty$.
\end{enumerate}
We say that the entry ($'0'$ or $'1'$) in the position $j$ at time $t$ has 
index $\phi_t(j)$.
From this definition, we notice that $n_1$ entries take 
the largest index $L_1$.
Furthermore, if we denote the number of entries of index $k$ by $I_k$, 
then $I_k$ does not change in time and is given as
\begin{equation}
I_k=
\left\{
\begin{array}{ll}
\ell_0 &\mbox{for} \; k=-\infty \\ 
n_1 &\mbox{for} \; L_2+1 \le |k| \le L_1 \\
n_1+n_2 &\mbox{for} \; L_3+1 \le |k| \le L_2 \\
\cdots & \qquad \\
\sum_{j=1}^s n_j &\mbox{for} \; 1 \le |k| \le L_s
\end{array}
\right.
\end{equation} 
For example, the indices of the sequence 
\begin{center}
\begin{tabular}{*{16}{@{\ \ }c}}        
0 & 0 & 0 & 1 & 1&  1 & 0 & 0 & 1 & 1&  0 & 1 & 1&  0&  0 & 0
\end{tabular}
\end{center}
are given as
\begin{center}
\begin{tabular}{*{16}{@{\ \ }c}}
-4&-$\infty$ &-$\infty$ & 4 & 2 & 1 & -1 & -2 & 3 & 1 & -1 & 2 & 1 & -1 &
 -2 & -3
\end{tabular}
\end{center}
Using the index, we define {\it solitons} and their position.
\begin{Definition}
A soliton consists of $'1'$s in a $01$ sequence
and can be determined by the following process.
\begin{enumerate}
\item Choose one of the $'1'$s whose index is $L_1$.
We suppose its position is $j_{L_1}$.
\item Choose the $'1'$ which is the nearest $'1'$ with index $L_1-1$ 
in the counterclockwise direction from the position $j_{L_1}$.
({\it i.e.} to the ``right" of that position.)
Let $j_{L_1-1}$ be its position.
\item Similarly, let $j_{L_1-2}$ be the position of its (counterclockwise) 
nearest $'1'$ with index $L_1-2$.
\item Repeat the procedure until the $'1'$ at $j_1$ is determined.
\item Then we define the set of $'1'$s at $j_{L_1}, j_{L_1-1}, \cdots, j_1$
to constitute a soliton with length $L_1$ and position $j_1$.
\item Perform the same procedure starting from one of the other 
$n_1-1$ $'1'$s with 
index $L_1$, and repeat it until all the $n_1$ solitons with length 
$L_1$ have been determined.
\item The largest index of the remaining $'1'$s is $L_2$.
Repeat the above 
procedure to the remaining $'1'$s and determine all 
$n_2$ solitons with length $L_2$.
\item In a similar manner, we determine $n_3$ solitons with length $L_3$, 
$n_4$ solitons with length $L_4$, ..., and finally we determine 
$n_s$ solitons with length $L_s$.
\end{enumerate}
\end{Definition}
\begin{figure}[htbp]
  \begin{center}
     \scalebox{0.7}{\includegraphics{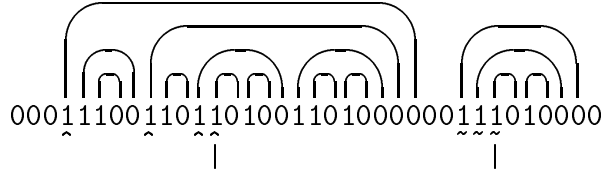}}
  \end{center}
  \vspace{-5mm}
  \caption{Solitons in a pBBS.}
  \label{fig:Solitons}
\end{figure}
In Fig.~\ref{fig:Solitons}, there are eight solitons (length $4$ $\cdots$ $1$,
length $3$ $\cdots$ $1$, length $2$ $\cdots$ $2$, and length $1$ $\cdots$ $4$).
The $'1'$s with $'\wedge '$ at the bottom constitute the 
largest soliton and those
with $'\sim'$ constitute the second largest soliton.
Their position is marked by $'|'$.
\par
Next, we define the notion of {\it a block}.
Recall that we drew arc lines between $10$ pairs to determine 
the time evolution of pBBS.
As is shown in Fig.~\ref{fig:Solitons}, a state of the pBBS is divided 
into disjoint $01$ sequences of such arc lines.
We call each disjoint sequence {\it a block}.
For example, there are two blocks in Fig.~\ref{fig:Solitons}.
The following properties of a block are obvious from its definition.
\begin{enumerate}
\item The numbers of $'0'$s and $'1'$s in a block are the same.
\item There is only one soliton with the largest length in each block.
\item After time evolution $t \to t+1$, all the $'0'$s in a block at $t$, 
turn into $'1'$s, and all the $'1'$s into $'0'$s. 
\item If the length of the largest soliton in a block is $L_M$, the $'1'$ 
at the left edge of the block has index $L_M$ and the $'0'$ at the right edge 
of the block has index $-L_M$.
\item The length of a block is twice the sum of the lengths of solitons 
contained in the block.
\end{enumerate}

The following Proposition, which plays an important role in subsequent
 proofs is a direct consequence of the 
definition of soliton and block.
\begin{Proposition}
\label{SolitonicNature}
Suppose that $'1'$s at the positions $j_L, j_{L-1}, ..., j_1$ constitute 
a soliton with length $L$, and that the $'0'$s which form the $10$ pairs 
with these $'1'$s are at the positions $i_1, i_2, ..., i_L$. 
The $'1'$ at $j_k$ has index $k$ and the $'0'$ at $i_k$ has index $-k$.
 Then,  
\begin{enumerate}
\item If an entry is located between $j_{k+1}$ and $j_k$, the absolute value of
 its index is less than $k$.
\item If an entry is located between $i_{k}$ and $i_{k+1}$, the absolute value 
of its index is less than or equal to $k$.
\item If we eliminate all the $'1'$s at $j_L, j_{L-1}, ..., j_1$ and all 
$'0'$s at $i_1, i_2, ..., i_L$, the entries with positions inbetween
 $j_{k+1}$ and $j_k$ and 
$i_k$ and $i_{k+1}$ $(k=1,2,...,L)$ constitute disjoint blocks. 
\item Let $\Delta(k)$ be the difference of the number of $'1'$s and $'0'$s
contained between $j_L$ and $k$, then $0 \le \Delta(k) \le L$.
In particular, $\Delta(k)=0$ iff $k=i_L$ and $\Delta(K) \le L-1$ for $k < j_1$.
Similarly, let $\Delta'(k)$ be the difference of the number of $'0'$s and 
 $'1'$s contained between $k$ and $i_L$, then $0 \le \Delta'(k) \le L$.
In particular, $\Delta'(k)=0$ iff $k=j_L$.
\end{enumerate}
\end{Proposition}

\begin{figure}[htbp]
  \begin{center}
     \scalebox{0.7}{\includegraphics{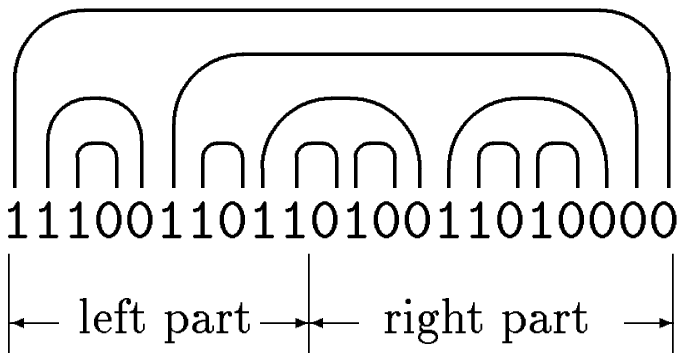}}
  \end{center}
  \vspace{-5mm}
  \caption{An example of a block and its left and right parts.}
  \label{fig:Block}
\end{figure}

Let us discuss some important properties of such a block.
We assume that the length of the largest soliton in the block is $L$, 
and that it 
is constituted by $'1'$s at positions $j_L, j_{L-1}, ..., j_1$, 
where $'1'$ at 
$j_k$ forms a $10$ pair with $'0'$ at $i_k$ ($1 \le k \le L$).
First we divide a block into two parts. (Fig.~\ref{fig:Block})
The {\it right part} of a block is the $01$ sequence which is located 
on the right side (counterclockwise) from the position of 
the largest soliton in the block.
In other words, the right part is the sequence from $i_1$ to $i_L$.
The remainder is called the {\it left part} of the block.
It is the sequence from $j_{L}$ to $j_1$.
The following properties are obvious.
\begin{enumerate}
\item The number of $'1'$s in the left part is greater by $L$ than that of 
$'0'$s.
On the contrary, the number of $'1'$s in the right part is less by $L$ than 
that of $'0'$s.
\item The entries at the edges of the left part are $'1'$ and those 
of the right part are $'0'$.
\end{enumerate}
Furthermore, from the rule for making $10$ pairs, we have the following 
Proposition.
\begin{Proposition}
\label{01sequence}
\begin{enumerate}
\item In any sequence starting from 
the left edge of the left part, the number of $'1'$s 
is greater than that of $'0'$s.
\item In any sequence ending at 
the right edge of the left part, the number of $'1'$s 
is greater than that of $'0'$s.
\item In any sequence ending at 
the right edge of the right part, the number of $'0'$s 
is greater than that of $'1'$s.
\item In any sequence starting from 
the left edge of the right part, the number of $'0'$s 
is greater than or equal to that of $'1'$s.
\end{enumerate}
\end{Proposition}
\Proof \hspace{12pt}
The statements 1 and 3 are obvious.
For the statement 2, we assume that the sequence starts from position $k$
and that $j_{m+1} \le k \le j_m$.
From Proposition \ref{SolitonicNature}, there can only be solitons with length
 less than or equal to $m-1$ inbetween $j_{m+1}$ and $j_m$.
From Proposition \ref{SolitonicNature}-2 and -3, the difference of 
the number of $'0'$s and $'1'$s between $k$ and $j_m-1$ is less 
than or equal to $m-1$.
On the other hand, there are $m$ more $'1'$s than $'0'$s between
 $j_{m}$ and $j_1$.
Hence statement 2 holds.
The statement 4 is proved in the similar manner.

\qed

Next we discuss some features of the time evolution of solitons.
A state of pBBS at time step $t-1$ is divided into blocks.
We pay attention to a single one of them.
The following Proposition is essential to understand the movement of solitons.
\begin{Proposition}
\label{BlockInvariance}
In a block, the number and type of solitons at $t-1$ do not change at $t$.
Furthermore, the position of the solitons at $t$ does not depend on the pattern
outside the block at $t-1$.
\end{Proposition}
The above Proposition is proved by induction using the Lemmas given below.
Let $L$ be the length of the largest soliton in the block.
The position of its components is denoted by $j_L, j_{L-1}, \ldots, j_1$ and that of the pairing $'0'$s by $i_1, i_2, \ldots, i_L$.
We refer to the sequence which, at $t$, is updated from 
the left (right) part of 
the block at $t-1$ as the left (right) part at $t$.
\begin{figure}[htbp]
  \begin{center}
     \scalebox{0.7}{\includegraphics{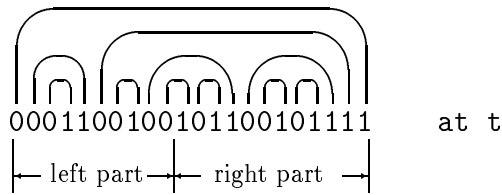}}
  \end{center}
  \vspace{-5mm}
  \caption{The left and right parts at $t$ updated from the block
 at $t-1$ shown in Fig.~\ref{fig:Block}.}
  \label{fig:Att}
\end{figure}
Fig.~\ref{fig:Att} shows the left and right parts at $t$ which are updated
from the state shown in Fig.~\ref{fig:Block}.
\begin{Lemma}
\label{Lemma4}
For $L \ge 3$, if $'10'$s are eliminated in the left part at $t$, 
there are at least three consecutive $'0'$s in the left edge.
If further $10$-elimination is possible and is performed on the sequence, 
there are at least four consecutive $'0'$s in the left edge.
In general, if $10$-elimination is performed $m$ times ($1 \le m \le L-2$),
there remain at least $m+2$ consecutive $'0'$s in the left edge.
\end{Lemma}
\Proof \hspace{12pt}
Recall that all the $'0'$s in a block at $t-1$ turn into $'1'$s, 
and all the $'1'$s into $'0'$s. 
From Proposition \ref{01sequence}-2 we have that every $'1'$ 
in the left part at $t$ has 
its pair $'0'$ within the left part.
Let $\Delta(k)$ be the difference of the number of $'0'$s and $'1'$s 
inbetween $j_L$ and $k$. 
From Proposition \ref{01sequence}-1 or \ref{SolitonicNature}-4
we find that
 $\DIS \min_{k \in [j_L, j_1]} \Delta(k) \ge 1$. 
Suppose that $\Delta_k=1$ at $k=k_i$ $\,(1\le i \le r)$, then the entry at $k_i$ is $'1'$ and the entry at $k_i+1$ is $'0'$. 
Hence, if we eliminate $'10'$s, the $'1'$s at $k_i$ are removed.
Let $J_2$ be the set of positions $J_2 = \{k| k \ne k_i, k_i+1\ (1\le i \le r),
\ k \in [j_L+1, j_1]\}$.
Since $\DIS \min_{k  \in J_2} \Delta(k) \ge 2$,there are at least three $'0'$s 
in the left edge of the reduced sequence.
To prove the Lemma, proceed in a similar manner.
\qed
\begin{Lemma}
\label{Lemma5}
Suppose that $L \ge 3$.
In the left part at $t$, if we perform $10$-elimination $L-2$ times, 
only $L$ consecutive $'0'$s remain and no $'1'$ is left.
\end{Lemma}
\Proof \hspace{12pt}
As in the proof of Lemma \ref{Lemma4}, we consider the difference $\Delta(k)$.
Since $10$-elimination always erases a $'1'$ and $'0'$ simultaneously, 
the value 
$\Delta(k)$ also indicates the difference of $'0'$ and $'1'$
in the reduced sequence, between its left edge $j_L$ and 
the entry at the (original) position $k$. 
From Proposition \ref{SolitonicNature}-4, $\Delta(k)$ ($j_L \le k \le j_1$) 
can only take its maximum value $L$ at $k=j_1$.  
On the other hand, from Lemma \ref{Lemma4}, at least $L$ consecutive 
$'0'$s remain after $L-2$ $10$-eliminations. 
Hence only $L$ consecutive $'0'$s
can remain in the left part at $t$. 
\qed

\begin{Lemma}
\label{Lemma6}
In the left part, the number and type of solitons at $t-1$ do not change at $t$
except for the largest soliton. Furthermore, the position of the solitons 
at $t$ does not depend on the pattern outside the block at $t-1$.
\end{Lemma}
\Proof
\hspace{12pt} 
If $L \le 2$, there is no $'1'$ in the left part at $t$ and
the above statement becomes trivial.
We therefore suppose $L \ge 3$.
Consider a state of a smaller pBBS of the same size as the left part 
in consideration.
We assume that the pattern of the state coincides with the left part
at $t-1$ except for the $'1'$s 
which constitute the largest soliton, which all are replaced with $'0'$s.
The number of $'0'$s in this state is greater than that of $'1'$s, by 
an amount of $L$.
If we update according to the time evolution rule, this pattern is exactly 
equal to that of the left part at $t$ (Proposition \ref{SolitonicNature}-3).
Since the number and type of solitons does not change in the sequence 
(Proposition \ref{conservation}) , to prove 
the Lemma, it suffices to show that no $'1'$ of the left part belongs to
a soliton whose position is outside that part 
and that no $'1'$ from the outside belongs to a 
soliton whose position is in the left part.

From Proposition \ref{01sequence}-2, $'1'$ does not form a pair with $'0'$ 
at the right edge (at the position $j_1$), which, at the same time, means
that no $'1'$ forms a pair with $'0'$ in the outside. 
(Remember that a pair means a $'1'$ and a $'0'$ which are 
connected by an arc line 
according to the time evolution rule at $t$.)
Since the pattern of the left part at $t$ is the same as the sequence 
considered above, the indices of $'1'$s coincide with those of that sequence.
Hence the largest index of the left part at $t$ is at most $L-2$.

There are $L$ more $'0'$s than $'1'$s in the left part at $t$, and the index of the $'0'$ at $j_1$ is less than or equal to $-L$.
Since the largest index in the left part is at most $L-2$, 
from Proposition \ref{SolitonicNature}-1, a $'1'$ in the left part 
does not belong to a soliton whose position is outside the left part.

There are at least two consecutive $'0'$s in the left edge of the left part
at $t$.
The index of the second $'0'$ from the left edge is at most $-2$, and,
by Proposition  \ref{SolitonicNature}-1, 
no $'1'$  in the outside belongs to a soliton whose position is 
in the left part if its index is less than or equal to $-2$.

Suppose that a $'1'$ of index $k$ ($k \ge 3$) outside the left part
constitutes a soliton with $'1'$s inside, whose indices 
are $k-1, k-2, \ldots,1$.
When we perform $10$-elimiantion $k-2$ times, from Lemma \ref{Lemma4}
we have that 
there are at least two consecutive $'0'$s between the inside $'1'$ with 
index $k-1$ and the outside $'1'$ of index $k$, which contradicts 
our definition of a soliton.

Thus we find that  no $'1'$ on the outside constitutes a soliton whose 
position is in the left part, which completes the proof.

\qed

As for the right part, we have the following Lemmas.
\begin{Lemma}
\label{Lemma7}
For $L \ge 2$, if $'10'$s are eliminated in the right part at $t$, 
there are at least three consecutive $'1'$s in the right edge.
If possible, further $10$-elimination is performed, 
there are at least four consecutive $'1'$s in the right edge.
In general, if $10$-elimination is performed $m$ times ($1 \le m \le L-2$),
there remain at least $m+2$ consecutive $'1'$s in the right edge.
\end{Lemma}
The proof is analogous to the proof of Lemma \ref{Lemma4}.
\begin{Lemma}
\label{Lemma8}
If we eliminate $'10'$s in the right part at $t$, $L-1$ times,
the reduced sequence consists of $L$ consecutive $'1'$s and no $'0'$.
\end{Lemma}
\Proof \hspace{12pt}
Let $J_m$ $(1 \le m \le L-2)$ be the set of positions of 
the entries in the right part at $t$
which remain after $m$ $10$-eliminations and that do not constitute
$m+2$ consecutive $'1'$s in the right edge. 
Then, as in the proof of Lemma \ref{Lemma4}, we have an inequality: 
$\min_{k \in J_m}\Delta'(k) =m+1$.
For $m=L-2$, this inequality implies that the sequence has the form
$$
\underbrace{1010 \cdots 10}_{\mbox{ repetition of $10$ }}
\hspace{-12pt}
\underbrace{11\cdots 1}_{\mbox{$L$}}
$$
Hence, by applying $10$-elimination $L-1$ times, 
we obtain a pattern with $L$ consecutive $'1'$s and no $'0'$.
\qed

Now we give the proof of Proposition \ref{BlockInvariance}
\Proof {\bf \,of Proposition \ref{BlockInvariance}}
\hspace{12pt}
From Proposition \ref{01sequence}-4,  any sequence starting from 
the left edge of the right part at $t$ contains 
a number of $'1'$s, larger than or equal to that of $'0'$s. 
(Recall that $'0'$ and $'1'$  in a block at $t-1$ turn 
to $'1'$ and $'0'$ at $t$.)
Hence a $'0'$ in the right part at $t$ forms a pair with one of 
the $'1'$s in the right part, and the pair is not affected 
by the outside pattern.
There are just $L$ $'1'$s which do not form pairs with inside $'0'$s.
These $'1'$s are those obtained by $10$-eliminations as in Lemma 
\ref{Lemma8}.
If their indices at $t$ are $L, L-1, \ldots,2,1$, then since the index of 
 $'0'$ at $j_1$ is at most $-L$ and all the indices in the right part are less 
than or equal to $L$ (Lemma \ref{Lemma8}), 
no $'1'$ outside the right block 
can belong to the inside solitons (Proposition \ref{SolitonicNature}-1) 
and the $L$ $'1'$s constitute a soliton with length $L$ whose position 
does not depend on the outside pattern.
In fact, its position is always $i_L$.
Then, together with Lemma \ref{Lemma6}, the solitons in the block 
do not depend on the outside pattern.
In other words, the positions of solitons at $t$, exactly coincide
with those in the case where no other block exists in the state.
According to Proposition \ref{conservation}, 
the solitons are preserved in time 
evolution and the Proposition is proved. 
Therefore we now only have to show that the indices of the  $L$ 
consecutive $'1'$s, 
which are obtained by $10$-eliminations in the right block at $t$,
 are $L, L-1, \ldots,2,1$.

By the definition of a block, the entry at $i_L+1$ is necessarily $'0'$, and
the $'1'$ at the right edge $i_L$ forms a pair with it.
Hence its index is $1$ and the statement is true for $L=1$.

When we eliminate all the $'10'$s in the state, the right edge of 
the ``reduced" block is $'1'$ (Lemma \ref{Lemma7}).
Since all the blocks, for which the largest soliton was of length $1$, 
are eliminated, 
and since the other blocks have at least one $'0'$ in the left edge, 
this $'1'$ forms a pair with a $'0'$ and its index is $2$ and the statement
 is true for $L \le 2$.
Similarly, when we eliminate the $'10'$s once more, the right edge of 
the reduced block is still $'1'$ (Lemma \ref{Lemma7}). 
Since all the blocks with the largest soliton whose length $L \le 2$ are 
eliminated, Lemma \ref{Lemma4} assures that the next right entry 
of the $'1'$ is $3$.

Repeating this argument, we find that the consecutive $'1'$s have 
indices $L, L-1, \ldots,1$, which completes the proof.
\qed

In the above proof, we have also shown that the position of the largest 
soliton  at $t$ is  the right edge of the block.
From Proposition  \ref{SolitonicNature}-3 and the fact that the length of 
the block is twice the sum of the lengths of solitons in the block, we  
obtain the following {Corollary}
\begin{Corollary}
\label{ScatteringLength}
The position of the largest soliton is updated to the right edge
of the block.
The difference between the position of the largest soliton at $t$
and that at $t-1$ is $L+2 \times \mbox{(sum of the lengths of solitons 
in the right part)}$.
\end{Corollary}

Since the $'0'$s in the right part at $t$ form pairs with the $'1'$s
 in the right part, we have
\begin{Proposition}
\label{RightPartCoinsidence}
The pattern of the right part at $t-1$ coincides with the pattern 
in the same region at $t+1$.
\end{Proposition}

Finally the movement of solitons is expressed by the Theorem:
\begin{Theorem}
\label{SolitonMovement}
Suppose that a pBBS has $M$ solitons. 
Let $X_j(t-1)$ and $L_j$ $(1 \le j \le M)$  be the positions and lengths of the solitons at $t-1$.
Then, their positions at $t$, $X_j(t)$ $(1 \le j \le M)$, satisfy
\begin{equation}
\label{SolitonFormula}
X_j(t)-X_j(t-1) \equiv L_j + \sum_{i=1}^{M} 2 x_j^i(t) \min[L_j,L_i] \qquad 
\mbox{\rm modulo $N$}, 
\end{equation}
where the soliton at $X_j(t)$ has length $L_j$ and the coefficients $x_j^i(t)$ 
($1 \le i, j \le M$) satisfy $x_j^i(t)=-x_i^j(t) \in \{-1,0,1\}$.
In particular, if $X_s(t)$ is the position of the largest soliton, $x_s^i(t) = 0$ or $1$.
\end{Theorem}
\Proof \hspace{12pt}
We prove the theorem by induction on the number of solitons $M$.
For $M=1$, the statement is trivial ($x_1^1(t)=0$).
Let $k \ge 1$ and assume that the theorem is true for $M \le k$.
When there are $k+1$ solitons, if the state at $t-1$ consists of 
multiple blocks, the latter part of Proposition
\ref{BlockInvariance} can be applied to each block and the
theorem is proved.
When the state consists of one block, as shown in the proof of 
Lemma \ref{Lemma6}, except for the largest soliton the other solitons 
in the left part are updated as if the largest soliton did not exist.
Hence they satisfy (\ref{SolitonFormula}) by the induction hypothesis.
Let $X_s(t-1)$ be the position of the largest soliton at $t-1$. 
From Corollary \ref{ScatteringLength},
\begin{eqnarray}
X_s(t)-X_s(t-1)&=
&L_s+2 \sum_{j\in\{ \mbox{\scriptsize $X_j(t-1)$ belongs to the right part} \}} L_j
\nonumber \\
&=&L_s+\sum_{j}2 x_s^j(t) \min[L_s,L_j],
\label{MoveOfLargest}
\end{eqnarray}
where $x_s^j(t)=1$ if the $j$th soliton is in the right part 
and otherwise $x_j^s(t)=0$.
Hence the largest soliton satisfies (\ref{SolitonFormula}).

On the other hand, by Proposition \ref{RightPartCoinsidence}, 
the solitons in the right part at $t$ are supposed to move to their positions
at $t-1$ if there is no soliton outside the region.
By the induction hypothesis this movement satisfies (\ref{SolitonFormula}).
Then, if the $j$th soliton is in the right part at $t$,
${}^{\exists} {x'}_j^i \in \{-1,0,1\}$,
\begin{equation}
X_j(t-1)-X_j(t) \equiv L_j + \sum_{i=1,i\ne s}^{M} 2 {x'}_j^i \min[L_j,L_i], 
\quad 
\mbox{modulo $N$}
\end{equation}
where ${x'}_j^i=-{x'}_i^j$ and ${x'}_j^i=0$ when the $i$th or $j$th soliton is 
outside the right part.
Since  $x_s^j(t)=1$ for the $j$th soliton in the right part, if we put
$x_j^s(t)=-1$, $x_j^i(t)=-{x'}_i^j$ $(i, j \ne s)$, then 
\begin{equation}
X_j(t)-X_j(t-1) \equiv L_j + \sum_{i=1}^{M} 2 {x}_j^i(t) \min[L_j,L_i]. \quad 
\mbox{modulo $N$}
\end{equation}
Thus the Theorem also holds true for $M=k+1$, which completes the proof.
\qed
So far, we have not obtained concrete expressions for $x_i^j(t)$,
however, Theorem \ref{SolitonMovement} and  Corollary \ref{ScatteringLength} 
are enough to determine the fundamental cycle of the pBBS for a given 
initial state. 

In the subsequent section, we shall use some properties of solitons 
with length $1$.
Hereafter we often refer to such a soliton as  a {\it $1$-soliton}.
The following Proposition is easily obtained from Proposition 
\ref{SolitonicNature}
\begin{Proposition}
\label{1-soliton-pattern}
Let $p$ be the position of a 1-soliton.
The entry at $p+1$ is necessarily a $'0'$.
If the entry at $p-1$ is $'1'$, then its index is greater than
$2$.
On the contrary, if the entries at $p-1, p, p+1$ are $'010'$, then
$p$ is the position of a 1-soliton.
Similarly if the entries at $p-1, p, p+1$ are $'110'$ and the
index of $'1'$ at $p-1$ is greater than two, then 
$p$ is the position of a 1-soliton.
In particular, if the entries from $p-1$ to $p+3$ are $'11011'$,
then $p$ is the position of a 1-soliton.
\end{Proposition}
As for the movement of the 1-soliton, we have
\begin{Proposition}
\label{Move-1-Soliton}
The position of a 1-soliton is updated by $+1$ or $-1$.
\end{Proposition}
\Proof \hspace{12pt} 
First we assume that the distance between the 1-solitons 
is at least $3$.
Let $p$ be the position of a $1$-soliton at $t$.
The entry at $p+1$ must be $'0'$.
If the entry at $p+2$ is $'1'$, then the entries from $p$ to $p+2$ 
form a pattern $'010'$ at $t+1$ and we see that the position changes 
by $+1$. 

When the entry at $p+2$ is $'0'$, then the entry at $p-1$ is $'0'$
by Proposition \ref{1-soliton-pattern}.
The index of $'0'$ at $p+2$ is either $-\infty$ or
$-k$ $(k \ge 2)$. If it is $-\infty$, the entries from $p$ to $p+2$ 
become a $'010'$ pattern at $t+1$ and we have that 
the position changes by $+1$. 
If it is $-k$, the index of $'0'$ at $p-1$, $\ell_{p-1}(t)$, satisfies 
$-k+1 \le \ell_{p-1}(t) \le -1$, hence the entries from $p-2$ to $p+2$
are updated as either $'01011'$ or $'11011'$.
By Proposition \ref{1-soliton-pattern}, the position of the soliton
is $p-1$ and hence it changes by $-1$.

When the distance between the 1-solitons is just $2$,
they form a sequence like 
$\DIS ...\underbrace{1010\cdots 10}_{\mbox{\rm $m$ times}}...$.
Let $p$ be the position of the leftmost 1-soliton.
Then we repeat the same arguments as above and find that all the
solitons move by $+1$ when the entry at $p+2m$ is $'1'$ or $'0'$
with index $-\infty$ and by $-1$ when it is $'0'$ with index $-k$.
\qed

%
%

\section{Fundamental cycle of the pBBS}
 In this section, we consider the fundamental cycle (the shortest period)
of a pBBS for a given initial state.
A key idea to determine the fundamental cycle is to establish 
a relation between the pBBS and the one obtained from it 
by $10$-elimiantion.
For a given state of the pBBS, we obtain a pattern of the state of a 
smaller pBBS by $10$-elimination.
A soliton with length $L$ ($L \ge 2$) turns into a soliton with length $L-1$,
and a 1-soliton disappears.
However, it is convenient to think of it as turning into a {\it $0$-soliton} 
--a soliton with length $0$-- which has no entry but has a {\it position}.
To make the idea clear, consider the following example.
A state
$$
000111000\mbox{\underline{$1$}}000011\mbox{\underline{$1$}}011000000
$$
contains two 1-solitons, the $'1'$s which are underlined. 
By $10$-elimination, we obtain a new pattern
$$
000110000011100000.
$$  
In this new sequence, however, we can denote the places where $1$-solitons 
existed before the 10-elimination as
$$
0001100|00011|100000.
$$  
So we may think that $0$-solitons exist
between two consecutive entries
separated by a $'|'$.
Of course, the notion of a {\it $0$-soliton} only has meaning when we eliminate
$1$-solitons  by $10$-elimination.
An advantage to consider the $0$-solitons is that we can devise a reverse 
operation for $10$-elimination.
I.e. we can reconstruct the original state from the new sequence if
we allow the existence of $0$-solitons.

When there are solitons with the same length, we can define the {\it
effective distance} between two of them with respect to the notion of 
a $0$-soliton.
\begin{Definition}
\label{effective-distance}
When there are two solitons with the same length, we perform $10$-elimination
repeatedly and turn them into $0$-solitons.
We denote the distance between these two $0$-solitons by the effective distance between them.
\end{Definition}
Here the distance between the two $0$-solitons is understood as the number
of entries between them.
For example, the effective distance between the above two $1$-solitons is
five.
An important property of the effective distance is
\begin{Proposition}
\label{Conservation-Distance}
The effective distance between two solitons does not 
change in time.
\end{Proposition}
\Proof \hspace{12pt}
By Lemma \ref{Lemma2}, $10$-elimination commutes with 
the updating rule of a pBBS. 
Hence it is enough to prove the Proposition for two $1$-solitons.

If there is no soliton with length more than one, the Proposition is trivial.
If the $1$-solitons themselves form blocks, {\it i.e.}, if they do not belong 
to a block together with a larger soliton, the statement of the Proposition 
is also obvious because they will move by $+1$ and the number of $'10'$s 
between the two solitons does not change at the next time step.

At time $t$, let us consider the block to which one of the $1$-solitons belongs.
The block is composed of consecutive $'1'$s and $'0'$s and they change
to $'0'$s and $'1'$s at the next time step.
Let $p_l$ and $p_r$, respectively, be the position of the 
$'1'$ at the left edge 
of the block and of the $'0'$ at the right edge.
Suppose that there are $m_r$ $'10'$s on the right of the $1$-soliton 
in the block, and $m_l$ $'10'$s on the left of it.
(Hence there are $m_r+m_l+1$ $'10'$s in the block.)
When we eliminate $'10'$s at $t$ the number of entries in the block
decreases by $2(m_r+m_l+1)$.
Let the distance between the $0$-soliton and the right (left) edge 
of the reduced block be $L_r$ ($L_l$).

From Proposition \ref{Move-1-Soliton}, the $1$-soliton moves by $+1$ 
or $-1$ in the next time step.
If it moves by $+1$, the number of $'10'$s in the block which exist 
to the right of the $1$-soliton does not change.
(We take into account the $'10'$ at the right edge of the block as well.)
Since the total number of $'10'$s does not change in a block at $t$,
the number of $'10'$s to the left of the $1$-soliton does not change either.
On the other hand, when it moves by $-1$, the number of $'10'$s to its
right increases by one and the number of those on its left decreases by one.
Hence, if we eliminate $'10'$s at $t+1$, 
the distance between the $0$-soliton and
the $'1'$ next to the right edge of the block, 
which was originally at the position $p_r-1$, 
is also $L_r$ in both cases.
It is also clear that the distance between the $0$-soliton and
the $'0'$ next to the left edge, which was at the position $p_l+1$, 
 is also $L_l$.
Therefore, when we perform $10$-elimination, the position of the $0$-soliton
in the reduced block does not change by the updating rule, where the 
reduced block at $t+1$ is defined as the sequence starting from the $'0'$ which was
at $p_l+1$ to the $'1'$ which was at $p_r-1$.
Since the distance between the reduced blocks does not change under 
time evolution, 
even for a block of $0$-soliton obtained from a block of 
$1$-solitons, the effective distance between two $1$-solitons does not change in
time evolution, which completes the proof.
\qed

From the above Proposition, we know that the effective distance is also 
a conserved quantity of the pBBS. 

The pBBS sometimes has some hidden symmetry which makes it difficult 
to determine the fundamental cycle.
The symmetry appears when there are solitons with the same length.
Suppose that there are more than one solitons with length $L$.
By applying $10$-elimination for $L$ times, 
these solitons turn into $0$-solitons.
Let $N_L$ be the size of the reduced pBBS. 
We denote 
the positions of the $0$-solitons in this reduced pBBS by
$\DIS p_1, p_2, \ldots, p_k \in \Z_{N_L}$.
Here $p_i$ is understood as the position of the right box adjacent to
the $i$th $0$-soliton.
Let $s_L$ be the smallest positive integer which satisfies
$$
\left\{p_1, p_2, \ldots,p_k  \right\}
= \left\{p_1+s_L, p_2+s_L, \ldots,p_k+s_L  \right\}. 
$$
(Note that the positions are defined modulo $N_L$.)
The number $s_L$ ($1 \le s_L \le N_L$) is a divisor of $N_L$
and we put $\DIS m_L:=N_L/s_L \in \Z$.
By Proposition \ref{Conservation-Distance}, $m_L$ does not change 
in time.
In this situation,
 we define the effective symmetry as follows.
\begin{Definition}
\label{Effective-Symmetry}
Let $m_L$ be the positive integer defined above.
Then, the solitons with length $L$ in the pBBS are said to 
have effective translational symmetry
of order $m_L$.
\end{Definition}
For example, a sequence 
$$
110000100011101001100010
$$
has three $1$-solitons. 
By $10$-elimination, we have
$$
1000|0011|0100|,
$$
{\it i.e.}; $N_1=15$, $s_1=5$ and hence the $1$-solitons have effective 
translational symmetry of order $3$.

 Now we fix the system and its initial state at $t=0$.
As in section 1, we consider a pBBS with $N$ boxes.
We assume that the initial state has $n_j$ solitons with length $L_j$ 
$\,(j=1,2, \ldots, s)$, where $L_1 > L_2 > \cdots > L_s$. 
We also put
$\ell_0 := N-\sum_{j=1}^l 2p_j = N-\sum_{j=1}^s 2 n_j L_j$, and
\begin{equation}
\ell_j := L_j-L_{j+1} \qquad (j=1,2,...,s-1).
\label{Ell_j}
\end{equation}
The initial state turns to a state of a smaller pBBS by $10$-elimination.
By {\it $k$-th pBBS}, we denote the pBBS which has as an initial state,
the one obtained by $L_{k+1}$ times $10$-elimination performed
on the original initial state.
The position of the origin of the $k$-th pBBS can be determined
 arbitrarily. 
The size of the $k$-th pBBS, $N_k$, is given as
\begin{equation}
N_k:= \ell_0+2 n_1 (L_1-L_{k+1}) + 2 n_2 (L_2-L_{k+1})+...+2 n_k(L_k-L_{k+1}).
\label{N_j}
\end{equation}
In particular, $N_0=\ell_0$.

The pBBS is then updated according to the time evolution rule 
and after some time period 
it  will take the same pattern as the initial state
except for some translation.
We call such period as a {\it relative cycle}.
The shortest relative cycle is called {\it the fundamental relative cycle}.
Apparently, the fundamental cycle is a relative cycle and an integer 
multiple of the fundamental relative cycle.

The following Lemma is used to establish the relation between the fundamental
cycle of the $k$-th pBBS and the fundamental relative cycle of the $k+1$-th
pBBS.
\begin{Lemma}
\label{Lemma9}
When the smallest soliton has length $\ell \ge 2$ in a pBBS,
the fundamental relative cycle of the pBBS coincides with that of 
the reduced pBBS obtained by $10$-elimination.
\end{Lemma}
\Proof \hspace{12pt}
From Lemma \ref{Lemma2}, we have that the fundamental relative cycle of 
the pBBS is the relative cycle of the reduced pBBS. 
Since the number of solitons does not change by $10$-elimination in this
case, we can reconstruct the pattern from the reduced pattern by adding $'10'$s
to the left of the solitons.
Thus the fundamental relative cycle of the reduced pBBS is also a relative 
cycle of the original pBBS.
\qed

For a pattern in a pBBS which contains at $t$ a $1$-soliton,
we obtain a pattern for a reduced pBBS by $10$-elimination.
The $1$-soliton turns into a $0$-soliton in that pattern.
If we now shift this pattern such that the position of this $0$-soliton 
is situated at the left (the midpoint between the position $0$ 
and $1$), we can define a map from a sequence of the pBBS to a
sequence of the reduced pBBS.
We shall denote this map by $\Psi(p(t))$.
Here $p(t)$ denotes the position of the $1$-soliton.
For example, a sequence
$$
0001110001\mbox{\underline{$10$}}110000
$$
is mapped by $\Psi(11)$ to
$$
|100000011001
$$
where $'|'$ denotes the position of $0$-soliton.

An important property of this map is that it commutes with
time evolution, namely,
\begin{Lemma}
\label{Lemma10}
Let $\mathcal{S}_t$ be a sequence of a pBBS at $t$ with a $1$-soliton at position $p(t)$. Then we have
\begin{equation}
\label{commuting.property}
\Psi(p(t+1))\mathcal{S}_{t+1}= \hat{T}\left(\Psi(p(t))\mathcal{S}_t \right),
\end{equation}
where $\hat{T}$ denote the time evolution in the reduced pBBS.
\end{Lemma}
\Proof \hspace{12pt}
By Lemma \ref{Lemma2}, the sequence $\Psi(p(t+1))\mathcal{S}_{t+1}$
is equivalent to $\hat{T}\left(\Psi(p(t))\mathcal{S}_t \right)$.
Hence it suffices to show that the position of one specific soliton coincides 
in both sequences.

The sequence $\mathcal{S}_t$ consisits of blocks.
First we consider the case 
where the $1$-soliton at $p(t)$ itself constitutes a block.
Let $L_l$ be the distance between the $1$-soliton at $t$ and the right edge of 
the block nearest to it on the left.
Since the position of the largest soliton in the block becomes 
the right edge of the block (Corollary \ref{ScatteringLength}),  
after $10$-elimination, there are $L_l-1$ $'0'$s between 
the $0$-soliton and the nearest soliton on its left. Namely, the position of
the soliton nearest to it on the left, is $L_l-2$.
On the other hand, the blocks of the sequence $\Psi(p(t))\mathcal{S}_t$
are obtained by  
eliminating all the $'10'$s, changing indices $k$ to $k-1$ 
and $-k$ to $-k+1$ $(k\ge 2)$, and letting the position of 
the $0$-soliton be the midpoint between the entries at position $0$ and $1$.
Since the updating rule is equivalent to changing $'0'$s in the block to $'1'$s
and $'1'$s to $'0'$, we find that the position of the largest soliton in
the block becomes $L_l-2$ at $t+1$.
Thus, we find that relation (\ref{commuting.property}) holds.

Next we consider the case where the $1$-soliton belongs to a block 
together with a larger soliton.
As is shown in the proof of Proposition \ref{Conservation-Distance},
the distance (number of entries) between the $0$-soliton and the right 
edge of the block at $t$, coincides with that between 
the position of the largest soliton and the $0$-soliton at $t+1$.
Since the distance from the $0$-soliton to the right edge 
is equal to the position of the largest soliton of the block in 
$\hat{T}\left(\Psi(p(t))\mathcal{S}_t \right)$,
the position of the largest soliton in the block in 
$\Psi(p(t+1))\mathcal{S}_{t+1}$ coincides with that in
$\hat{T}\left(\Psi(p(t))\mathcal{S}_t \right)$, which completes the proof.
\qed

From this Lemma, and  because the effective 
distance between $1$-solitons does not change 
by Proposition \ref{Conservation-Distance}, we find that 
the positions of all the $0$-solitons in $\Psi(p(t))\mathcal{S}_t$
coincide with those in $\Psi(p(t+1))\mathcal{S}_{t+1}$. We state this fact as a Lemma:
\begin{Lemma}
\label{Lemma11}
The positions of all the $0$-solitons in $\Psi(p(t))\mathcal{S}_t$
coincide with those in $\Psi(p(t+1))\mathcal{S}_{t+1}$
\end{Lemma}

Using these Lemmas, we can prove the key Proposition:
\begin{Proposition}
\label{Fundamental-Relation}
The fundamental cycle of the $k$-th pBBS is a relative cycle of
the $k+1$-th pBBS. If the smallest solitons in the $k+1$-th pBBS do not have
effective translational symmetry, it is the fundamental relative 
cycle of the $k+1$-th pBBS.
\end{Proposition}
\Proof \hspace{12pt}
We denote the pattern at $t$ of the $k+1$-th pBBS by $\mathcal{S}_t$.
We choose one of the $1$-soliton and denote its position at $t$ by $p(t)$.
Without loss of generality, we can assume the sequence of the $k$-th system
at $t=0$ to coincide with $\Psi(p(0))\mathcal{S}_0$.
By Lemma \ref{Lemma10},  we have
$$
\Psi(p(t+1))\mathcal{S}_{t+1}= \hat{T}\left(\Psi(p(t))\mathcal{S}_t \right),
$$
where $\hat{T}$ is the time evolution operator for the $k$-th pBBS.
Therefore we find 
\begin{equation}
\Psi(p(t))\mathcal{S}_t = \hat{T}^t\Psi(p(0))\mathcal{S}_0.
\label{commutativity}
\end{equation}
If $T$ is the fundamental cycle of the $k$-th pBBS,
(\ref{commutativity}) implies
$$
\Psi(p(T))\mathcal{S}_T = \Psi(p(0))\mathcal{S}_0.
$$
Since by Lemma \ref{Lemma11} the above relation holds up 
to the position of $0$-solitons,
$\mathcal{S}_T$ is equivalent to $\mathcal{S}_0$.
Therefore $T$ is a relative cycle of the $k+1$-th pBBS.
Conversely, if $T$ is the fundamental relative cycle of the $k+1$-th pBBS,
$\mathcal{S}_T$ is equivalent to $\mathcal{S}_0$, and ${}^\exists p,\ 
 \Psi(p(T))\mathcal{S}_T = \Psi(p) \mathcal{S}_0$. 
However, if the $1$-solitons in the $k+1$-th pBBS do not have
the effective translational symmetry, $p$ must be $p(0)$ by Lemma 
\ref{Lemma11}. Thus we find
$$
\hat{T}^T\Psi(p(0))\mathcal{S}_0 = \Psi(p(0))\mathcal{S}_0,
$$
which means that $T$ is a cycle of the $k$-th pBBS.
Since $T$ is the shortest relative cycle of the $k+1$-th pBBS
and the fundamental cycle of $k$-th pBBS is a relative cycle 
of the $k+1$-th pBBS, $T$ is the fundamental cycle of the 
$k$-th pBBS.
\qed

Now we determine the fundamental cycle of the pBBS for a given initial
state.
We divide the initial states into two cases.
\hbreak
{\large \bf Case 1}  {\it Solitons without effective translational symmetry}
\hbreak
First we assume that no soliton has effective translational symmetry.
Since there are only solitons with length $\ell_1$ and the size of the
system is $N_1$, the fundamental cycle of the first system $t_1$ is obtained as
\begin{equation}
t_1=\frac{\LCM(N_1,\ell_1)}{\ell_1},
\label{t_1}
\end{equation}
where $\LCM(N_1, \ell_1)$ denotes the least common multiple of $N_1$ and $\ell_1$.
During this period, a soliton passes the origin (the position of the
 $0$-soliton) 
of the first pBBS exactly $r_1$ times, where
\begin{eqnarray}
r_1&= &\frac{t_1 \ell_1}{N_1} \nonumber \\
&= &\frac{\LCM(N_1,\ell_1)}{N_1}.
\label{r_1}
\end{eqnarray}

From Proposition \ref{Fundamental-Relation}, the fundamental relative cycle
of the second pBBS is $t_1$. 
The fundamental cycle $T_2$ is an integer multiple of the relative cycle.
To determine $T_2$, we look for the distance $s_2$ 
over which the largest soliton 
will move during the period $t_1$.
Since the largest soliton in the second pBBS has 
length $\ell_1+\ell_2$ and it collides $r_1$ times with
each one of the smaller solitons 
during the fundamental relative cycle,
by Corollary \ref{ScatteringLength}, we obtain
\begin{equation}
\label{s_2}
s_2=t_1 (\ell_1+\ell_2)+2 r_1 n_2 \ell_2.
\end{equation}
Hence, the fundamental cycle of the second system is given as $T_2 = t_1 t_2$, 
where
\begin{equation}
t_2 = \frac{\LCM(N_2,s_2)}{s_2}.
\label{t_2}
\end{equation}
During this period $T_2$, the largest soliton passes  $r_2$ times
through the origin of the second pBBS, where
\begin{eqnarray}
r_2&= &\frac{t_2 s_2}{N_2} \nonumber \\
&= &\frac{\LCM(N_2,s_2)}{N_2}.
\label{r_2}
\end{eqnarray}
In a similar manner, the largest soliton in the third pBBS has length $\ell_1+\ell_2+\ell_3$ and its shift during the period $T_2$
is given as
\begin{equation}
s_3 = t_1 t_2 (\ell_2+\ell_2+\ell_3)+2 r_1 n_2 t_2 (\ell_2+\ell_3)+2 r_2 n_3 \ell_3.
\label{s_3}
\end{equation}
The fundamental cycle $T_3$ is given as $T_3 =t_1 t_2 t_3$, where
\begin{equation}
t_3=\frac{\LCM(N_3,s_3)}{s_3},
\label{t_3}
\end{equation}
and it passes the origin $r_3$ times, where
\begin{equation}
r_3=\frac{\LCM(N_3, s_3)}{N_3}.
\label{r_3}
\end{equation} 
Repeating the above procedure, we find that the fundamental cycle of
the pBBS is given by the following theorem
\begin{Theorem}
\label{MainTheorem1}
When there is no effective translational symmetry of solitons in a pBBS,
the fundamental cycle $T$ of the pBBS is given as
\begin{equation}
T=\prod_{k=1}^s t_k,
\label{Fundamental-Cycle}
\end{equation}
where $t_k$ is obtained recursively as
\begin{eqnarray}
s_1&=&\ell_1\\
t_1&=&\frac{\LCM(N_1,s_1)}{s_1}\\
r_1&=&\frac{\LCM(N_1,s_1)}{N_1}\\
s_k&=&t_1t_2\cdots t_{k-1}(\ell_1+\ell_2+...+\ell_k)\nonumber \\
   &&\quad + 2r_1n_2t_2t_3\cdots t_{k-1}(\ell_2+\ell_3+...+\ell_k)
    \nonumber\\
&&\quad +2r_2n_3t_3\cdots t_{k-1}(\ell_3+\ell_4+...+\ell_k)+\ldots + 2 r_{k-1}n_k \ell_k\\
t_k&=&\frac{\LCM(N_k,s_k)}{s_k}\\
r_k&=&\frac{\LCM(N_k,s_k)}{N_k}
\label{Recursion}
\end{eqnarray}
Here $\ell_k$ and $N_k$ $(k=1,2,...,s)$ are defined as in (\ref{Ell_j}) 
and (\ref{N_j}).
\end{Theorem}
In order to rewrite this  formula in a compact way, we note first that
\begin{equation}
s_k=r_{k-1}N_k + \frac{t_1 t_2 \cdots t_{k-1} \ell_k \ell_0}{N_{k-1}}.
\label{s_k-N_k-relation}
\end{equation}
Equation (\ref{s_k-N_k-relation}) is proved by induction.
We give a proof in Appendix B.
Noticing the identity
$$
\frac{\LCM(s+jN,N)}{s+jN} =\frac{\LCM(s,N)}{s} \qquad {}^{\forall} j \in \Z
$$
we have
\begin{equation}
t_k=\frac{\LCM(N_k, q_k)}{q_k},
\label{New-t_k}
\end{equation}
where $q_1=\ell_1$ and 
\begin{equation}
q_k:=\frac{t_1 t_2 \cdots t_{k-1}\ell_k \ell_0}{N_{k-1}} \qquad (k \ge 2).
\label{q_k}
\end{equation}
The map $\LCM$ $(\Z_{+} \times \Z_{+} \mapto \Z_{+})$ is naturally extended 
to the map $\Q_{+} \times \Q_{+} \mapto \Q_{+}$ as
\begin{eqnarray}
&&\LCM\left( 2^{a_1}3^{a_2}5^{a_3}7^{a_4}\cdots, 2^{b_1}3^{b_2}5^{b_3}7^{b_4}\cdots \right) \nonumber \\
&&\qquad =  2^{\max[a_1,b_1]} 3^{\max[a_2,b_2]} 5^{\max[a_3,b_3]} 7^{\max[a_4,b_4]}\cdots \quad (a_j, b_j \in \Z).
\label{ExtensionLCM}
\end{eqnarray}
We also define $\DIS \LCM(a,b,c):= \LCM\left( \LCM(a,b) ,c \right)$ {\it etc.},
as $\DIS \LCM\left( \LCM(a,b) ,c \right)=\LCM\left( a, \LCM(b,c) \right)$.
From (\ref{New-t_k}) and (\ref{q_k}) we obtain,
$$
t_k=\LCM\left(\frac{N_{k-1} N_k}{t_1 t_2 \cdots t_{k-1} \ell_k \ell_0},1 \right)
$$
and
$$
\prod_{j=1}^k t_j =\LCM\left(\frac{N_{k-1} N_k}{\ell_k \ell_0}, t_1 t_2 \cdots t_{k-1} \right).
$$
Thus we have 
\begin{eqnarray*}
\prod_{j=1}^s t_j &= &\LCM\left(\frac{N_{s-1} N_s}{\ell_s \ell_0}, t_1 t_2 \cdots t_{s-1} \right)\\
&=&\LCM\left(\frac{N_{s-1} N_s}{\ell_s \ell_0}, \frac{N_{s-2} N_{s-1}}{\ell_{s-1} \ell_0}, t_1 t_2 \cdots t_{s-2}\right)\\
&=& \cdots .
\end{eqnarray*}
Hence, from (\ref{Fundamental-Cycle}), we obtain
\begin{Corollary}
\label{FundamentalCorollary}
\begin{equation}
T=\LCM \left(
\frac{N_s N_{s-1}}{\ell_s \ell_0}, \frac{N_{s-1} N_{s-2}}{\ell_{s-1} \ell_0},
\cdots, \frac{N_1 N_0}{\ell_1 \ell_0}, 1
\right).
\label{MainFormula}
\end{equation}
\end{Corollary}
For example, the pBBS with the initial state
$$
00111011100100011110001101000000
$$
has four types of solitons with lengths 5, 4, 2, and 1.
The values of the parameters are caluculated as
 $\ell_1=1$, $\ell_2=2$, $\ell_3=1$, $\ell_4=1$, 
$N_0\equiv\ell_0=4$, $N_1=6$, $N_2=14$, $N_3=20$ and $N_4=32$.
Hence its fundamental cycle is
\begin{eqnarray*}
T &= & \LCM \left( \frac{32\cdot 20}{1\cdot 4},
\frac{20\cdot 14}{1\cdot 4},
\frac{14\cdot 6}{2\cdot 4},
\frac{6\cdot 4}{1\cdot 4}, 1
\right) \\
&= &\LCM\left( 160, 70, \frac{21}{2}, 6, 1 \right) \\
&= &2^5\cdot 3\cdot 5 \cdot 7  \\
&= &3360. \\
\end{eqnarray*}

\noindent
{\large \bf Case 2}  {\it Solitons with effective translational symmetry}
\hbreak
The main difference with the previous case is that the fundamental 
cycle of the 
$k$-th pBBS is a relative cycle of the $k+1$-th pBBS,
but not necessarily the fundamental relative cycle.
When the largest solitons do not have effective symmetry
however, the fundamental relative cycle can be determined easily
and the above arguments require only slight modification.
In the first pBBS the $0$-solitons, obtained by $10$-elimination on the
second pBBS, are arranged periodically. 
If we put $\DIS N_1^{*} := \frac{N_1}{m_2}$, $N_1^{*}$ is the period
of the translational symmetry.
Hence the fundamental relative cycle of the second pBBS $t_1^{*}$ is given as
\begin{equation}
\label{Symt_1}
t_1^{*}= \frac{\LCM(N_1^{*},\ell_1)}{\ell_1}.
\end{equation}
If we put
\begin{equation}
r_1^{*}= \frac{\LCM(N_1^{*},\ell_1)}{N_1^{*}},
\label{Symr_1}
\end{equation}
a largest soliton collides with $0$-solitons 
exactly $\DIS \frac{r_1^{*} n_2}{m_2}$ times
during the period $t_1^{*}$. Hence, in the second pBBS, the largest soliton
moves the distance
\begin{equation}
\label{Syms_2}
s_2^{*} = t_1^{*} (\ell_1+\ell_2) + 2 r_1^{*} n_2^{*} \ell_2
\end{equation}
where $\DIS n_2^{*} = \frac{n_2}{m_2}$.
Similarly, the fundamental relative cycle of the third pBBS is
\begin{equation}
t_2^{*}=\frac{\LCM(N_2^{*},s_2^{*})}{s_2^{*}}.
\label{Symt_2}
\end{equation}
We can proceed in a similar manner to the above and 
find that all the formulae given in Theorem \ref{MainTheorem1}
 hold with the change $N_j \to N_j^{*}$, $t_j \to t_j^{*}$ {\it etc.},
that is, 
the fundamental cycle $T$ of the pBBS is given as
\begin{equation}
T=\prod_{k=1}^s t_k^{*},
\label{F-Cycle2}
\end{equation}
where $t_k^*$ is obtained recursively as
\begin{eqnarray}
s_1^{*}&=&\ell_1\\
t_1^{*}&=&\frac{\LCM(N_1^{*},s_1^{*})}{s_1^{*}}\\
r_1^{*}&=&\frac{\LCM(N_1^{*},s_1^{*})}{N_1^{*}}\\
s_k^{*}&=&t_1^{*}t_2^{*}\cdots t_{k-1}^{*}(\ell_1+\ell_2+...+\ell_k)
\nonumber \\
&&\quad+2r_1^{*}n_2^{*}t_2^{*}t_3^{*}\cdots t_{k-1}^{*}
(\ell_2+\ell_3+...+\ell_k)\nonumber \\
&&\qquad +2r_2^{*}n_3^{*}t_3^{*}\cdots t_{k-1}^{*}(\ell_3+\ell_4+...+\ell_k)+\ldots + 2 r_{k-1}^{*}n_k^{*} \ell_k\\
r_k^{*}&=&\frac{\LCM(N_k^{*},s_k^{*})}{N_k^{*}}\\
t_k^{*}&=&\frac{\LCM(N_k^{*},s_k^{*})}{s_k^{*}}.
\label{Recursion2}
\end{eqnarray}
Here $\DIS N_k^{*}=\frac{N_k}{m_{k+1}}$,
$\DIS n_k^{*}=\frac{n_k}{m_{k}}$, 
$m_k$ is the order of the effective translational symmetry of the
$k$th largest solitons for $1 \le k \le s$, $m_{s+1}=1$, and
$\ell_k$ and $N_k$ $(k=1,2,...,s)$ are defined as in (\ref{Ell_j}) 
and (\ref{N_j}).

However, as in this case, there is no relation like 
(\ref{s_k-N_k-relation}),
we so far have not found a compact expression like (\ref{MainFormula}).

When the largest soliton has effective translational symmetry, 
the determination of the fundamental cycle becomes a little complicated.
For example, the sequence of the first pBBS 
$$
|1100|0011|0000 \qquad \mbox{at $t$}
$$
is updated as
$$
|0011|0000|1100 \qquad \mbox{at $t+1$},
$$
where $'|'$ denotes the $0$-solitons obtained from the second pBBS.
Apparently, the fundamental relative cycle of the second pBBS
is just one time step.
Although the largest soliton moves only two boxes, the distance 
between a pattern at $t$ and the same pattern at $t+1$ is eight.
Hence, in this example, we have to think of the largest soliton
as being shifted by a distance of eight boxes.
In general, when we shift the largest solitons by $\DIS N_1^{*}=\frac{
N_1}{\LCM(m_1,m_2)}$, the sequences coincide up to translation.
Here $m_k$ denotes the order of the effective translational symmetry
of the $k$th largest solitons.
Then the fundamental relative cycle of the second pBBS is given as
\begin{equation}
t_1^{*}=\frac{\LCM (N_1^{*}, \ell_1)}{\ell_1}.
\label{Newet_1}
\end{equation}
The distance separating it from the same pattern, $d$, 
is calculated as follows.
Let $(p_1, q_1) \in \Z \times \Z$ be a solution to the Diophantine equation
\begin{equation}
t_1^{*} \ell_1 + \frac{N_1}{m_1} p_1=\frac{N_1}{m_2}q_1.
\label{Diophantine}
\end{equation}
Then 
$$
d=\frac{N_1}{m_2}q_1 (=t_1^{*} \ell_1 + \frac{N_1}{m_1} p_1)
\qquad \mbox{modulo $N_1$}.
$$
In this expression, $q_1$ is the number of $0$-solitons between a 
largest soliton at $t$ and the `corresponding' largest soliton at $t+t_1^{*}$;
$\DIS p_1 n_1^{*}$ is the number of largest solitons
between the largest soliton at $t+t_1^{*}$ 
which `really' moved during the time
period $t_1^{*}$ and the corresponding largest soliton at $t+t_1^{*}$. 
Hence, in the second pBBS, the shift of the largest soliton is considered to be
\begin{equation}
s_2^{*}=t_1^{*} (\ell_1+\ell_2) + 2(p_1 n_1^{*} + q_1 n_2^{*})\ell_2 + p_1 \frac{N_1}{m_1}.
\label{News_2}
\end{equation}
When $k_2 :=\GCD(m_1,m_2)\ne 1$, the second pBBS also 
has translational symmetry and the fundamental relative cycle of the
third pBBS becomes $t_1^{*} t_2^{*}$, where
\begin{eqnarray}
t_2^{*}&= &\frac{\LCM( N_2^{*}, s_2^{*})}{s_2^{*}}, 
\label{Newet_2}\\
N_2^{*} &= &\frac{N_2}{\LCM(k_2,m_3)}.
\label{NewN_2}
\end{eqnarray}
The shift of the largest solitons in the third pBBS 
is obtained in a similar manner as
\begin{eqnarray}
s_3^{*}&= &t_1^{*}t_2^{*}(\ell_1+\ell_2+\ell_3) + t_2^{*}
\left( 2(p_1 n_1^{*} + q_1 n_2^{*})(\ell_2+\ell_3) + p_1 \frac{N_1}{m_1}
\right) \nonumber \\
&&\quad +2\left( \frac{n_1+n_2}{k_2} p_2 + n_3^{*}q_2  \right)\ell_3 + \frac{N_2}{k_2} p_2,
\label{News_3}
\end{eqnarray}
where $(p_2, q_2) \in \Z \times \Z$ is a solution to the Diophantine equation
\begin{equation}
t_2^{*} s_2^{*} + \frac{N_2}{k_2} p_2 =\frac{N_2}{m_3} q_2.
\label{Diophantine2}
\end{equation}
The fundamental relative cycle of the fourth pBBS is obtained 
from $s_3^{*}$ in a similar manner to (\ref{Newet_2}).
Repeating the same procedure, we obtain the fundamental cycle of the pBBS
as $T=\prod_{j=1}^s t_j^{*}$, where
\begin{eqnarray}
t_j^{*} &= &\frac{\LCM(N_j^{*}, s_j^{*})}{s_j^{*}} 
\label{Newt_j}\\
N_j^{*} &= &\frac{N_j}{\LCM(k_j,m_{j+1})} \qquad (m_{s+1}:=1) \\
k_j &= &\GCD(m_1, m_2, \cdots, m_j) \\
s_1^{*} &= &\ell_1   \\
s_j^{*} &= &\prod_{i=1}^{j-1}t_i^{*}\left(\sum_{i=1}^j \ell_j\right) 
            + \prod_{i=2}^{j-1}t_i^{*}\left[2\left(n_1^{*}p_1+n_2^{*}q_1\right)
       \left(\sum_{i=2}^j \ell_j\right) +\frac{N_1}{m_1}p_1 \right]\nonumber \\
           &&\quad +\prod_{i=3}^{j-1}t_i^{*}  \left[
                2\left(\frac{n_1+n_2}{k_2}p_2+n_3^{*}q_2\right)
          \left(\sum_{i=3}^j \ell_j\right) +\frac{N_2}{k_2} p_2 \right]\nonumber \\          && \qquad \quad \cdots \nonumber \\
           &&\qquad +2\left( \frac{\sum_{i=1}^{j-1} n_i}{k_{j-1}}p_{j-1}
            + n_j^{*} q_{j-1} \right)\ell_j + \frac{N_{j-1}}{k_{j-1}} p_{j-1}
\label{News_j}
\end{eqnarray}
and $(p_j, q_j) \in \Z \times \Z $ is a solution to the Diophantine equation
\begin{equation}
t_j^{*} s_j^{*} + \frac{N_j}{k_j} p_j = \frac{N_j}{m_{j+1}} q_j.
\label{DiophantineEq}
\end{equation}

Therefore we have found a formula to obtain the fundamental
cycle of the pBBS for an arbitrary initial state.
\begin{Theorem}
\label{NextTheorem1}
The fundamental cycle $T$ of the pBBS is given as
\begin{equation}
T=\prod_{j=1}^s t_j^{*},
\label{Fundamental-Cycle2}
\end{equation}
where $t_j^{*}$ is obtained recursively from (\ref{Recursion2}) 
if the largest solitons have no effective translational symmetry,
or from (\ref{Newt_j}) in the general case.
\end{Theorem}

\section{Concluding remarks}
In this article, we have shown a formula to determine the fundamental
cycle of a pBBS for a given initial state.
The pBBS is obtained from the ultradiscretization of 
the discrete Toda equation with periodic boundary condition.
Since the Toda equation with periodic boundary conditions has 
quasiperiodic solutions 
which are expressed by theta functions associated to
 hyperelliptic curves\cite{KvM,DT},
one may expect that the fundamental cycle has some relation
with the period matrices of the theta functions.
Establishing such a relation is a problem we wish to address in
the future.

Although we have treated the pBBS with box capacity one and
with only one kinds of balls, it can be extended to the case  with
box capacity larger than one and many kinds of balls.
The determination of the fundamental cycle for such an extended pBBS
is also a future problem.

Finally, we wish to point out one important property of the original BBS
which can be derived from the results in the previous sections.
Note that the BBS corresponds to the case of the pBBS where
 the number of boxes $N \to +\infty$.
The solitons are defined in exactly the same way and they are
also conserved in time.
From Corollary \ref{ScatteringLength}, Lemma \ref{Lemma10} and \ref{Lemma11},
we obtain
\begin{Proposition}
\label{Ninfty}
In the original BBS, after sufficient time steps,
the solitons are arranged according to
the order of their lengths and move freely.
\end{Proposition}
In the BBS, solitons as defined here behave like {\it solitons} in 
nonlinear differential equations, which justify the use of the 
word `soliton'.
\hbreak
\hbreak
\noindent
{\large  \bf Acknowledgements}

The authors are grateful to Prof. J. Satsuma and Dr. R. Willox 
for helpful comments on the present work.


\hbreak

\appendix{\Large \bf Appendix}
\section{Proof of Proposition \ref{StatesCount} }
\hbreak
We use the terminology of sections 2 and 3.
For a given state of the pBBS, we construct a state of the $k$-th pBBS
by performing $10$-elimination $L_{k+1}$ times.
In particular, after $L_1$ eliminations, we obtain a sequence of
$\ell_0$ consecutive $'0'$s.
We then fix the position of one of such $'0'$s.
To obtain a state of the first system, we should put $n_1$ $0$-solitons 
next to these consecutive $'0'$s.
The number of such possible configurations is 
$\DIS {\ell_0+n_1-1 \choose n_1}$.
Each configuration corresponds to a state of the first system.
To obtain a state of the second system, we put $n_2$ $0$-solitons 
next to the previous state.
The number of possible configurations is $\DIS {N_1+n_2-1 \choose n_2}$.
Repeating this procedure, we see that the number of states
for a fixed $'0'$ is 
$$
{\ell_0+n_1-1 \choose n_1}
{N_1+n_2-1 \choose n_2}{N_2+n_3-1 \choose n_3}\cdots 
{N_{s-1}+n_s-1 \choose n_s}.
$$
There are $N$ possibilities for the position of the fixed $'0'$, 
but any one of $\ell_0$ $'0'$s gives the same results.
Thus we find that the number of states $\Omega(N;\{L_j, n_j \})$ 
is given as
$$
\Omega(N;\{L_j, n_j \}) = \frac{N}{\ell_0} \prod_{i=1}^s
{N_{i-1}+n_i-1 \choose n_i}.
$$
\qed
\section{Proof of (\ref{s_k-N_k-relation})}
\hbreak
We prove (\ref{s_k-N_k-relation}) by induction.
Since $N_1=\ell_0+2 n_1 \ell_1$, $N_2=\ell_0+2n_1 \ell_1 +2(n_1+n_2) \ell_2$
and $r_1 N_1 = t_1 \ell_1$, we have
\begin{eqnarray*}
s_2 &= &t_1\left[ (\ell_1+\ell_2)+\frac{r_1}{t_1}\cdot 2n_2 \ell_2 \right]\\
&= &t_1\left[ (\ell_1+\ell_2)+\frac{\ell_1}{N_1}\cdot 2n_2 \ell_2 \right]\\
&= &\frac{t_1}{N_1}\left[ \ell_1\left( \ell_0+2n_1(\ell_1+\ell_2) 
+2 n_2 \ell_2 \right) +\ell_0 \ell_2  \right]\\
&= &r_1 N_2 +\frac{t_1\ell_2 \ell_0}{N_1}.
\end{eqnarray*}
Hence (\ref{s_k-N_k-relation}) is true for $k=2$.
Suppose that (\ref{s_k-N_k-relation}) holds up to $k=m-1$ $(m \ge 3)$.
We rewrite $s_m$ as
\begin{eqnarray}
s_m &= &t_1t_2 \cdots t_{m-1}( \ell_1 +\ell_2 + \cdots +\ell_m) + 
        2r_1t_2t_3\cdots t_{m-1}(\ell_2+\cdots +\ell_m) \nonumber\\
    &&\quad + \cdots +2r_{m-1}n_{m}\ell_m \nonumber\\
&= &t_{m-1}\Bigl[ s_{m-1}+\Bigl( (t_1t_2\cdots t_{m-2})+(2r_1
t_2 \cdots t_{m-2})n_2  \nonumber \\
&& \qquad +\cdots +2r_{m-2}n_{m-1}
\Bigr)\ell_m \Bigr] +2 r_{m-1} n_m \ell_m
\label{A0}
\end{eqnarray} 
Using the relation $\ r_k N_k = t_k s_k\  ({}^{\forall}k)$ and 
$N_m=N_{m-1}+\sum_{j=1}^m 2n_j \ell_m$, we have
\begin{eqnarray}
s_m &=&r_{m-1}N_m + t_{m-1}\ell_m \Bigl[    
t_1t_2\cdots t_{m-2} +2r_1 t_2\cdots t_{m-2}n_2 \nonumber\\
&& +\cdots +2r_{m-2}n_{m-1}
-\frac{2 s_{m-1}}{N_{m-1}}(\sum_{j=1}^{m-1}n_j)
\Bigr]
\label{A1}
\end{eqnarray}
By the induction hypothesis,
$$
s_{m-1}=r_{m-2}N_{m-1} + \frac{t_1 t_2 \cdots t_{m-2} \ell_{m-1} \ell_0}{N_{m-2}}
$$
and
\begin{eqnarray*}
&&2r_{m-2}n_{m-1}
-\frac{2 s_{m-1}}{N_{m-1}}(\sum_{j=1}^{m-1}n_j)\\
&&\quad=-\frac{2t_{m-1}s_{m-2}}{N_{m-2}}(\sum_{j=1}^{m-2} n_j) -t_1t_2
\cdots t_{m-2}
(\sum_{j=1}^{m-1} n_j ) \frac{\ell_{m-1}\ell_0}{N_{m-1}N_{m-2}}.
\end{eqnarray*}
Consecutive use of the induction hypothesis yields, 
\begin{eqnarray}
&&t_1t_2\cdots t_{m-2}+2r_1t_2 \cdots t_{m-2}n_2
 +\cdots +2r_{m-2}n_{m-1}-\frac{2 s_{m-1}}{N_{m-1}}(\sum_{j=1}^{m-1}n_j)\nonumber\\
&&=\left(\prod_{j=1}^{m-2} t_j\right) \ell_0 \left(
\frac{1}{N_1}-\frac{2\ell_2(n_1+n_2)}{N_1}{N_2}-\cdots -\frac{2 \ell_{m-1}(\sum_{j=1}^{m-1}n_j)}{N_{m-2}N_{m-1}}
\right)\nonumber\\
&&=\left(\prod_{j=1}^{m-2} t_j\right) \ell_0 \left( \frac{1}{N_2}-\frac{2\ell_3(\sum_{j=1}^3 n_j)}{N_2 N_3}-\cdots-\frac{2 \ell_{m-1}(\sum_{j=1}^{m-1}n_j)}{N_{m-2}N_{m-1}}
\right)\nonumber\\
&&=\qquad \cdots\nonumber\\
&&=\frac{t_1 t_2 \cdots t_{m-2} \ell_0}{N_{m-1}}
\label{A2}
\end{eqnarray}
Substituting (\ref{A2}) into (\ref{A1}), we obtain
\begin{equation}
s_m=r_{m-1}N_m + \frac{t_1 t_2 \cdots t_{m-1} \ell_m \ell_0}{N_{m-1}}.
\end{equation}
Therefore (\ref{s_k-N_k-relation}) is proved.
\qed
\end{document}